\documentclass[11pt,a4paper]{article} 
\pdfoutput=1

\usepackage{jcapmod} 
\usepackage{booktabs}
\usepackage{url}
\usepackage{setspace}

\newcommand{\eq}[1]{eq. (\ref{#1})} \newcommand{\fig}[1]{Figure \ref{#1}}
\newcommand{\sektion}[1]{section \ref{#1}}

\newcommand{\ocite}[1]{ref. \cite{#1}}

 \newcommand{\ud}{\mathrm{d}}

\newcommand{\bea}{\begin{eqnarray}}
\newcommand{\eea}{\end{eqnarray}}

\title{The Observational Status of Simple Inflationary Models: an Update}

\author[a]{Nobuchika Okada,}
\author[b]{Vedat Nefer \c{S}eno\u{g}uz}
\author[c]{and Qaisar Shafi}

 \affiliation[a]{Department of Physics and Astronomy,
University of Alabama, Tuscaloosa, AL 35487, USA}
\affiliation[b]{Department of Physics, Mimar Sinan Fine Arts University, 34380
\c{S}i\c{s}li, \.Istanbul, Turkey}
\affiliation[c]{Bartol Research Institute, Department of Physics and Astronomy, University of Delaware, Newark, DE 19716, USA}

\abstract{We provide an update on five relatively well motivated 
inflationary models in which
the inflaton is a Standard Model singlet scalar field. These include
i) the textbook quadratic and quartic potential models but with additional
couplings of the inflaton to fermions and bosons, which enable reheating and
also modify the naive predictions for the scalar spectral index $n_s$ and $r$, 
ii) models with Higgs and Coleman-Weinberg potentials, and finally 
iii) a quartic potential model with non-minimal coupling of the inflaton to gravity.
For $n_s$ values close to 0.96, as determined by
the WMAP9 and Planck experiments, most of the considered models predict $r\gtrsim0.02$.  
The running of the scalar spectral index, quantified by $|\ud
n_s/\ud\ln k|$, is predicted in these models to be of order
$10^{-4}$--$10^{-3}$.}

\begin{document} \maketitle \flushbottom

\section{Introduction} \label{intro}
The dramatic announcement of a B-mode polarization signal possibly due to
inflationary gravitational waves by the BICEP2 experiment \cite{Ade:2014xna}
brought new attention to  a class of inflationary models in which the energy
scale during inflation is on the order of $10^{16}$ GeV. Subsequent results by
the Planck experiment \cite{Adam:2014bub,Planck:2015xua} and the
joint Planck -- BICEP analysis \cite{Ade:2015tva} indicate that most (if not
all) of the signal observed by the BICEP experiment was caused by galactic dust.
However, a significant contribution from inflationary gravitational waves is not
ruled out. The joint Planck -- BICEP analysis provides a best fit value around
$0.05$ for the tensor to scalar ratio $r$. Although this result is not
statistically significant as it stands, it will soon be tested by forthcoming
data.

Motivated by these rapid developments in the observational front, in this paper we briefly review and update the
results of five closely related, well motivated and previously studied
inflationary models which are consistent with values of $r$ around 0.05, 
a signal level which will soon be probed.
The first two models employ the
very well known quadratic ($\phi^2$) and quartic ($\phi^4$) potentials
\cite{Linde:1983gd}, supplemented in our case by additional couplings of the
inflaton $\phi$ to fermions and/or scalars, so that reheating becomes
possible. These new interactions have previously been shown
\cite{NeferSenoguz:2008nn,Rehman:2010es,Martin:2014vha} to 
significantly modify the predictions for the
scalar spectral index $n_s$ and $r$ in the absence of these new interactions. 

The next two models exploit respectively
the Higgs potential
\cite{Rehman:2010es,Martin:2014vha,Destri:2007pv,Smith:2008pf,Okada:2013vxa}
and Coleman-Weinberg potential
\cite{Smith:2008pf,Rehman:2008qs,Shafi:1983bd,Shafi:2006cs}.
With the SM electroweak symmetry presumably broken by a Higgs potential, it
seems natural to think that nature may have utilized the latter (or the closely
related Coleman-Weinberg potential) to also implement inflation, albeit with a
SM singlet scalar field.

Finally, we consider
a class of models \cite{Okada:2010jf,Okada:2011en} which invokes a quartic
potential for the inflaton field, supplemented by an additional non-minimal
coupling of the inflaton field to gravity \cite{Martin:2014vha,Salopek:1988qh}.

Our results show that the predictions for $n_s$ and $r$ from these models
are generally in good agreement with the BICEP2, Planck and WMAP9 measurements,
except the radiatively corrected quartic potential which is ruled out
 by the current data. We display
the range of $r$ values allowed in these models that are consistent with $n_s$ being close to 0.96.  Finally, we present the predictions
for $|\ud n_s/\ud\ln k|$ which turn out to be of order $10^{-4}$--$10^{-3}$.

Before we discuss the models,
let's recall the basic equations used to calculate the inflationary parameters.
The slow-roll parameters may be defined as
(see \ocite{Lyth:2009zz}
for a review and references):
\begin{equation}
\epsilon =\frac{1}{2}\left( \frac{V^{\prime} }{V}\right) ^{2}\,, \quad
\eta = \frac{V^{\prime \prime} }{V}  \,, \quad
\zeta ^{2} = \frac{V^{\prime} V^{\prime \prime\prime} }{V^{2}}\,.
\end{equation}
Here and below we use units $m_P=2.4\times10^{18}\rm{~GeV}=1$, 
and primes denote derivatives with respect
to the inflaton field $\phi$.
The spectral index
$n_s$, the tensor to scalar ratio
$r$ and the running of the spectral index
$\alpha\equiv\mathrm{d} n_s/\mathrm{d} \ln k$ are given in the slow-roll
approximation by
\begin{equation}
n_s = 1 - 6 \epsilon + 2 \eta \,,\quad
r = 16 \epsilon \,,\quad
\alpha = 16 \epsilon \eta - 24 \epsilon^2 - 2 \zeta^2\,.
\end{equation}

The amplitude of the curvature perturbation $\Delta_\mathcal{R}$ is given by
\begin{equation} \label{perturb}
\Delta_\mathcal{R}=\frac{1}{2\sqrt{3}\pi}\frac{V^{3/2}}{|V^{\prime}|}\,,
\end{equation}
which should satisfy $\Delta_\mathcal{R}^2= 2.215\times10^{-9}$
from the Planck measurement \cite{Ade:2013zuv} with the pivot scale chosen at $k_0 = 0.05$
Mpc$^{-1}$.

The number of e-folds is given by
\begin{equation} \label{efold1}
N=\int^{\phi_0}_{\phi_e}\frac{V\rm{d}\phi}{V^{\prime}}\,, \end{equation}
where $\phi_0$ is the inflaton value at horizon exit of the scale corresponding
to $k_0$, and $\phi_e$ is the inflaton value at the end of inflation, defined by 
 max$(\epsilon(\phi_e) , |\eta(\phi_e)|,|\zeta^2(\phi_e)|) = 1$. The value of $N$
depends logarithmically on the energy scale during inflation as well as the reheating
temperature, and is typically around 50--60.

\section{Radiatively corrected quadratic and quartic potentials}
Inflation driven by scalar potentials of the type
\begin{equation}
V = \frac12 m^2 \phi^2 +\frac{\lambda}{4!} \phi^4
\end{equation}
provide a simple realization of an inflationary scenario
\cite{Linde:1983gd}. However, the inflaton field $\phi$ must have couplings to
`matter' fields which allow it to make the transition to hot big
bang cosmology at the end of inflation.
Couplings such as $(1/2)h \phi \bar{N} N$ or $(1/2)g^2 \phi^2 \chi^2$ (to a Majorana
fermion $N$ and a scalar $\chi$ respectively) induce correction terms to the potential 
which, to leading order, 
take the Coleman-Weinberg form \cite{Coleman:1973jx}
\begin{equation} \label{deltav}
 V_{\rm loop}\simeq -\kappa  \phi^4 \ln\left(\frac{\phi}{\mu}\right)\,.
\end{equation}
Here, $\mu$ is a renormalization scale which we set to $\mu=m_P$\footnote{For the radiatively
corrected quartic potential the observable inflationary parameters do not depend on the choice of the renormalization scale. However, this may not be the case for the radiatively corrected
quadratic potential, as discussed in \ocite{Enqvist:2013eua}.}, 
and $\kappa = (2h^4-g^4)/(32\pi^2)$ in the one loop approximation. 

\begin{figure}[!t]
\begin{center}
\scalebox{0.67}{\includegraphics{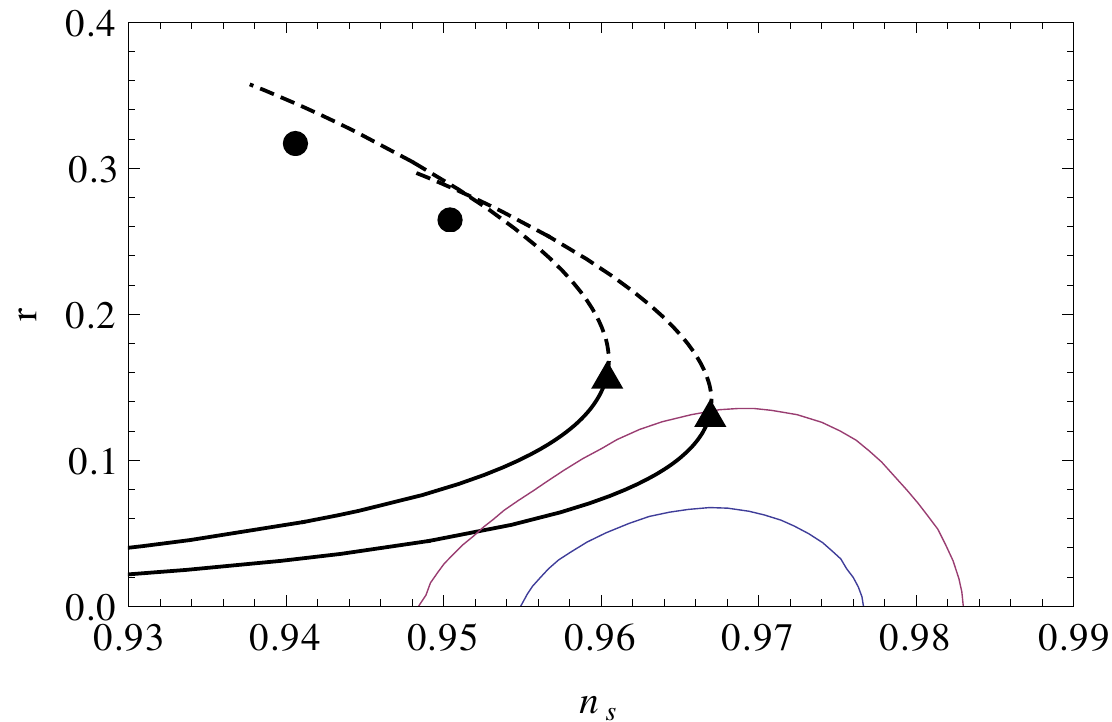}}
\scalebox{0.67}{\includegraphics{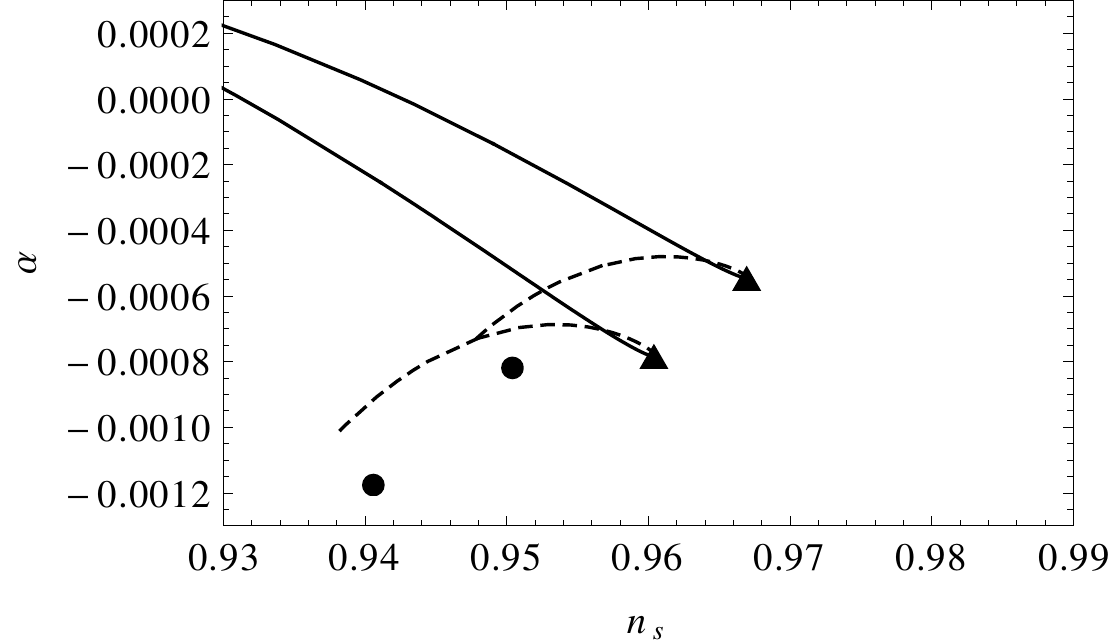}}
\end{center}\vspace{-0.5cm}
\caption{Radiatively corrected $\phi^2$ potential:
$n_s$ vs. $r$ (left panel) and $n_s$ vs. $\alpha$ (right panel) for various
$\kappa$ values, 
 along with the $n_s$ vs. $r$ contours (at the confidence levels of 68\% and 95\%) 
 given by the Planck collaboration (Planck TT+lowP) \cite{Planck:2015xua}. 
The black points and triangles are predictions 
 in the textbook quartic and quadratic potential models, respectively. The
dashed portions are for $\kappa<0$.  
$N$ is taken as 50 (left curves) and 60 (right curves). }
  \label{phi2fig}
\end{figure}

First, assume that $\lambda\ll m^2 / \phi^2$ during inflation, so that inflation is
primarily driven by the quadratic $\phi^2$ term. In the absence of radiative
corrections, this quadratic case of the well-known monomial model \cite{Linde:1983gd} predicts 
\begin{equation}
\label{quadratic}
n_s = 1-2/N\,,\quad r =8/N\,, \quad \alpha =-2/N^2\,.
\end{equation}

As discussed in
\ocite{NeferSenoguz:2008nn}, when $\kappa$ is positive there are two solutions
for a given $\kappa$. The ``$\phi^2$ solution'' approaches the tree level
result \eq{quadratic} as $\kappa$ decreases, whereas inflation takes place close
to the local maximum
for the ``hilltop solution'', resulting in a strongly tilted red spectrum with suppressed $r$.  
As the value of $\kappa$ is increased, the two branches of solutions
approach each other and they meet at $\kappa\simeq7\times10^{-15}$ for $N=60$.
For negative $\kappa$ values the $\phi^4\ln\phi$ correction term 
in the potential leads to predictions similar to those for the quartic potential given by
\begin{equation} \label{quartic}
n_s = 1-3/N\,,\quad r =16/N\,, \quad \alpha =-3/N^2\,.
\end{equation}

For each case, 
we calculate the inflationary predictions scanning over various
values of $\kappa$, while keeping the number of e-folds fixed.
Figure~\ref{phi2fig} shows the predictions for 
 $n_s$, $r$ and $\alpha$ with the number of e-folds 
 $N=50$ (left curves in each panel) and $N=60$ (right curves in each panel), 
 along with the Planck results \cite{Planck:2015xua}.
The values of parameters for selected values of $\kappa$ are displayed in
Table \ref{phi2tab}. 

\begin{figure}[!b]
\begin{center}
\scalebox{0.67}{\includegraphics{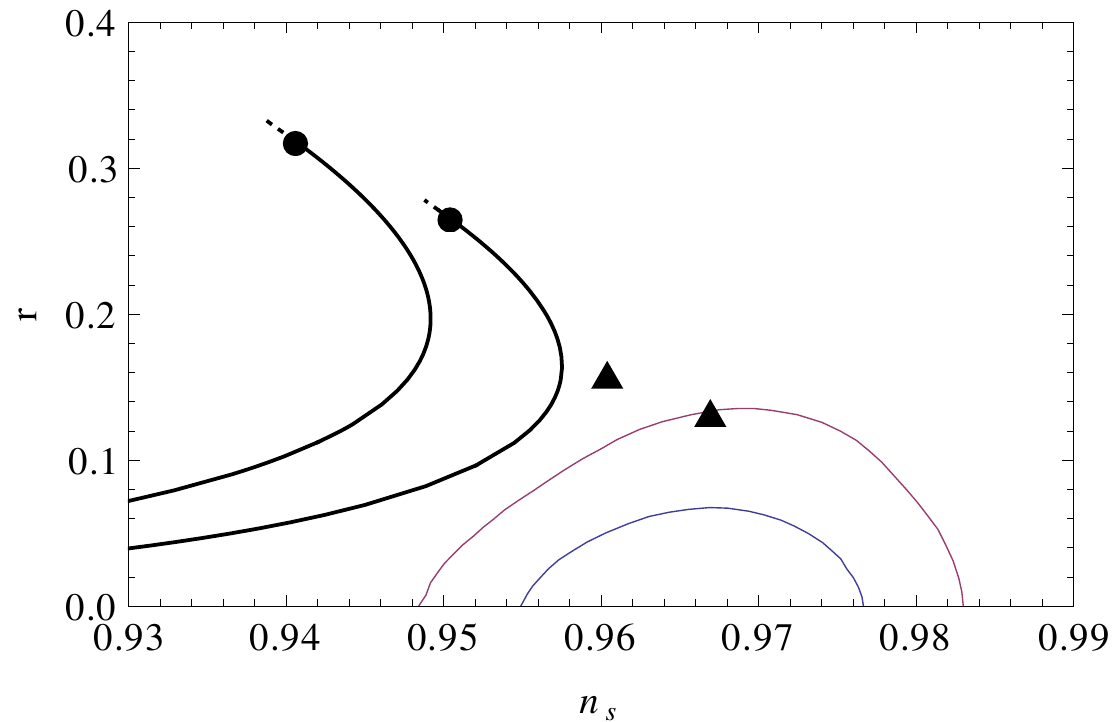}}
\scalebox{0.67}{\includegraphics{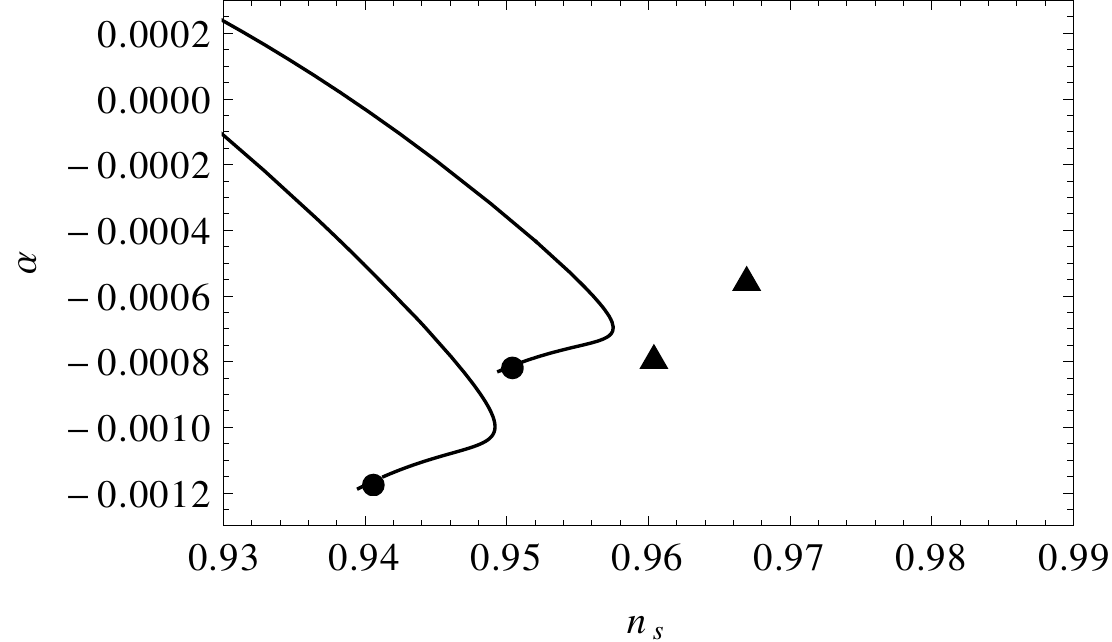}}
\end{center}\vspace{-0.5cm}
\caption{Radiatively corrected $\phi^4$ potential:
$n_s$ vs. $r$ (left panel) and $n_s$ vs. $\alpha$ (right panel) for various
$\kappa$ values, 
 along with the $n_s$ vs. $r$ contours (at the confidence levels of 68\% and 95\%) 
 given by the Planck collaboration (Planck TT+lowP) \cite{Planck:2015xua}. The black points and triangles are predictions 
 in the textbook quartic and quadratic potential models, respectively. The
dashed portions (extending just beyond the black points) are for $\kappa<0$. 
 $N$ is taken as 50 (left curves) and 60
(right curves).
}
  \label{phi4fig}
\end{figure}

The one loop contribution to $\lambda$ is of order $(4!)\kappa$, which is $\sim m^2 / \phi^2$
in the parameter range where the $\kappa$ term has a significant effect on
inflationary observables. In this case our assumption $\lambda\ll m^2 / \phi^2$
corresponds to the renormalized coupling being small compared to the one loop contribution.
Alternatively, assume that $\lambda\gg m^2 / \phi^2$ during inflation, so that inflation is
primarily driven by the quartic term. The numerical results for this case are
displayed in \fig{phi4fig} and  Table \ref{phi4tab}. 
As before, there are two solutions for a given
positive value of $\kappa$, and the predictions interpolate between a strongly
tilted red
spectrum with suppressed $r$ to the tree level result given in \eq{quartic}.
For negative $\kappa$ values the potential during inflation interpolates between
$\phi^4$ and $\phi^4\ln\phi$ potentials, as a consequence the predictions
remain close to \eq{quartic}.

\clearpage

\begin{table}[!t]
\begin{center}
\begin{tabular}{cccccccc}
\hline \hline
$\log_{10}(|\kappa|)$ & $m$ (GeV) & $V(\phi_0)^{1/4}$ (GeV) & $\phi_0$ & $\phi_e$ 
 & $n_s$  & $r$ & $\alpha\ (10^{-4})$    \\
\hline \hline
\multicolumn{8}{c}{negative $\kappa$ branch}\\\hline
 $-$14.0 & $1.38\times 10^{13}$ & $2.23\times 10^{16}$ & 17.0 & 1.42 & 0.962 & 0.215 & $-$4.81 \\\hline
 $-$14.5 & $1.46\times 10^{13}$ &$ 2.07\times 10^{16} $& 16.0 & 1.41 & 0.967 & 0.159 & $-$5.32 \\\hline
 $-$16.0 & $1.46\times 10^{13} $& $1.98\times 10^{16} $& 15.6 & 1.41 & 0.967 & 0.133 & $-$5.46 \\\hline
\hline
\multicolumn{8}{c}{$V=(1/2)m^2\phi^2$ }\\
\hline
& $1.46\times 10^{13}$ & $1.98\times 10^{16}$ & 15.6 & 1.41 & 0.967 & 0.132 &
$-$5.46\\
\hline \hline
\multicolumn{8}{c}{$\phi^2$ branch}\\\hline
 $-$16.0 & $1.46\times 10^{13}$ & $1.97\times 10^{16}$ & 15.5 & 1.41 & 0.967 & 0.131 & $-$5.47 \\\hline
 $-$14.5 & $1.41\times 10^{13}$ & $1.85\times 10^{16}$ & 15.0 & 1.41 & 0.965 & 0.102 & $-$5.15 \\\hline
 $-$14.3 & $1.30\times 10^{13}$ & $1.69\times 10^{16}$ & 14.4 & 1.41 & 0.959 & 0.070 & $-$3.79 \\\hline
 $-$14.2 & $1.22\times 10^{13}$ & $1.59\times 10^{16}$ & 14.0 & 1.41 & 0.954 &
0.056 & $-$2.59 \\\hline
\hline
\multicolumn{8}{c}{Hilltop branch}\\
\hline
 $-$14.2 & $1.01\times 10^{13}$ & $1.37\times 10^{16}$ & 13.2 & 1.41 & 0.940 &
0.031 &
0.58 \\\hline
 $-$14.3 & $7.9\times 10^{12}$ & $1.16\times 10^{16}$ & 12.5 & 1.41 & 0.921 &
0.016 & 3.41\\\hline
\hline
\end{tabular}
\end{center}
\caption{Radiatively corrected $\phi^2$ potential:
The values of parameters 
for number of e-folds $N=60$, in units $m_P=1$ 
unless otherwise stated. } 
\label{phi2tab}
\end{table}

\clearpage

\begin{table}[!t]
\begin{center}
\begin{tabular}{cccccccc}
\hline \hline
 $\log_{10}(|\kappa|)$ & $\log_{10}(\lambda)$ & $V(\phi_0)^{1/4}$ (GeV) & $\phi_0$ & $\phi_e$ 
 & $n_s$  & $r$ & $\alpha\ (10^{-4})$    \\
\hline \hline
\multicolumn{8}{c}{negative $\kappa$ branch}\\\hline
 $-$14.3 & $-$12.4 & $2.36\times 10^{16}$ & 22.6 & 3.67 & 0.950 & 0.269 & $-$8.11 \\\hline
 $-$15.0 & $-$12.1 & $2.34\times 10^{16}$ & 22.3 & 3.49 & 0.951 & 0.262 & $-$7.97 \\\hline
\hline
\multicolumn{8}{c}{$V=(1/4!)\lambda\phi^4$}\\\hline
& $-$12.1 & $2.34\times 10^{16}$ & 22.2 & 3.46 & 0.951 & 0.260 & $-$7.93\\\hline
\hline
\multicolumn{8}{c}{$\phi^4$ branch}\\\hline
 $-$15.0 & $-$12.0 & $2.34\times 10^{16}$ & 22.1 & 3.44 & 0.951 & 0.258 & $-$7.90 \\\hline
 $-$14.0 & $-$11.8 & $2.30\times 10^{16}$ & 21.5 & 3.28 & 0.953 & 0.241 & $-$7.64 \\\hline
 $-$13.5 & $-$11.5 & $2.17\times 10^{16}$ & 20.3 & 3.12 & 0.957 & 0.193 & $-$7.23 \\\hline
 $-$13.3 & $-$11.4 & $1.99\times 10^{16}$ & 19.1 & 3.03 & 0.957 & 0.135 & $-$6.24 \\\hline
 $-$13.26 & $-$11.3 & $1.87\times 10^{16}$ & 18.5 & 3.00 & 0.954 & 0.106 & $-$4.96\\\hline
\hline
\multicolumn{8}{c}{Hilltop branch}\\\hline
 $-$13.26 & $-$11.4 & $1.70\times 10^{16}$ & 17.8 & 2.97 & 0.947 & 0.073 & $-$2.28 \\\hline
 $-$13.30 & $-$11.4 & $1.55\times 10^{16}$ & 17.2 & 2.95 & 0.937 & 0.051 & 0.51 \\\hline
 $-$13.35 & $-$11.5 & $1.44\times 10^{16}$ & 16.8 & 2.94 & 0.929 & 0.038 & 2.61 \\\hline
\hline
\end{tabular}
\end{center}
\caption{Radiatively corrected $\phi^4$ potential:
The values of parameters  
for number of e-folds $N=60$, in units $m_P=1$ 
unless otherwise stated. } 
\label{phi4tab}
\end{table}

\section{Higgs potential} \label{higgs}

In this section we consider an inflationary scenario 
 with the potential of the form, 
\begin{eqnarray} 
V= \frac{\lambda}{4!} \left( 
 \phi^2 -v^2
\right)^2, 
\end{eqnarray}
where $\phi$ is the inflaton field, 
 $\lambda$ is a real, positive coupling, 
 and $v$ is the vacuum expectation value (VEV) 
 at the minimum of the potential. This Higgs potential was first
considered for inflation in ref. \cite{Vilenkin:1994pv}, and more
recently in refs. 
\cite{Martin:2014vha,Destri:2007pv,Smith:2008pf,Rehman:2008qs}.
Radiative corrections to the
Higgs potential were analyzed in refs. \cite{Rehman:2010es,Okada:2013vxa}.
Here, for simplicity, we have assumed that the inflaton is 
 a real field, but it is straightforward to extend the model 
 to the Higgs model, where the inflaton field is the Higgs field 
 and a gauge symmetry is broken by the inflaton VEV.

\begin{figure}[!t]
\begin{center}
\scalebox{0.4}{\includegraphics{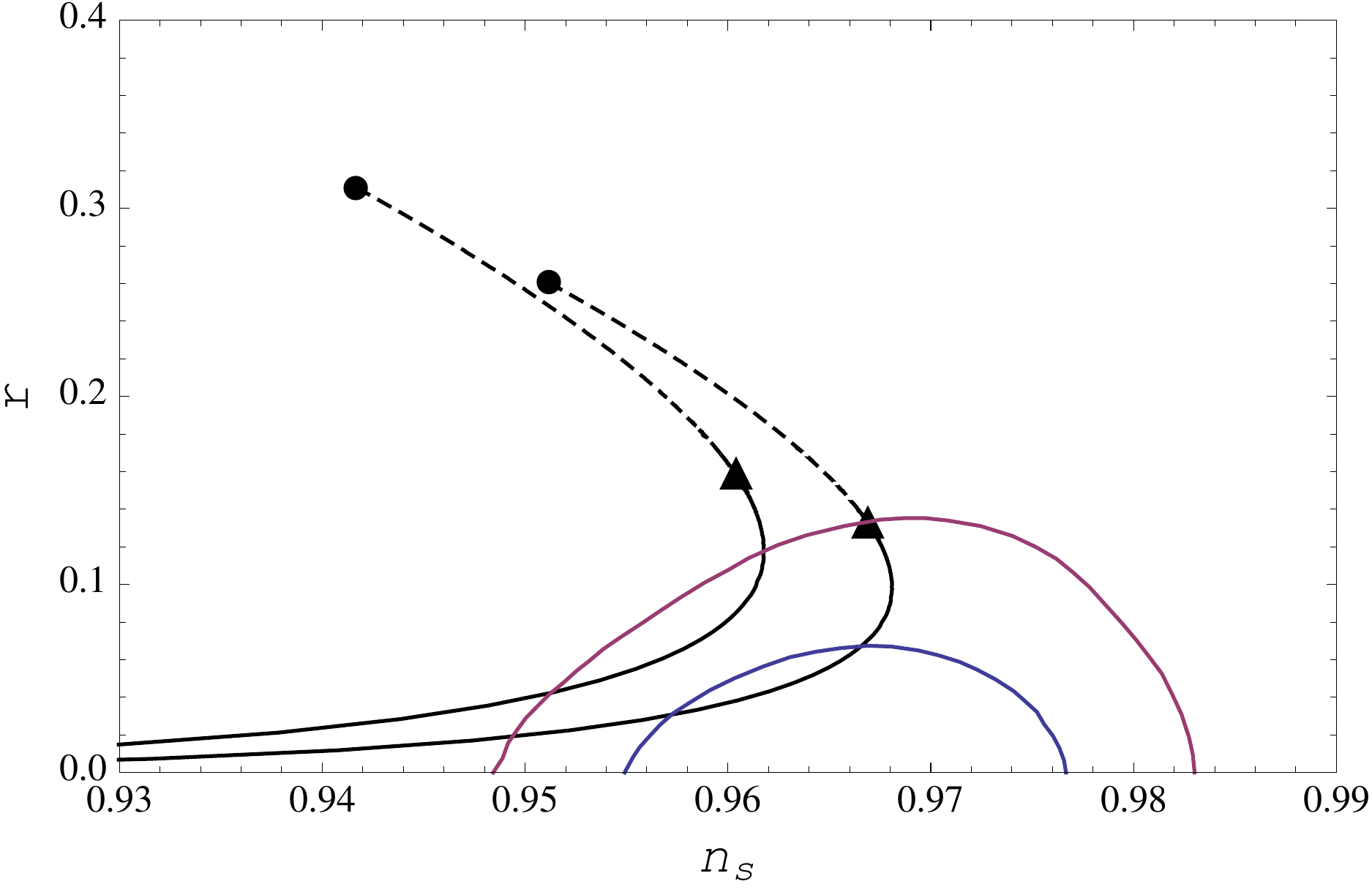}}
\scalebox{0.9}{\includegraphics{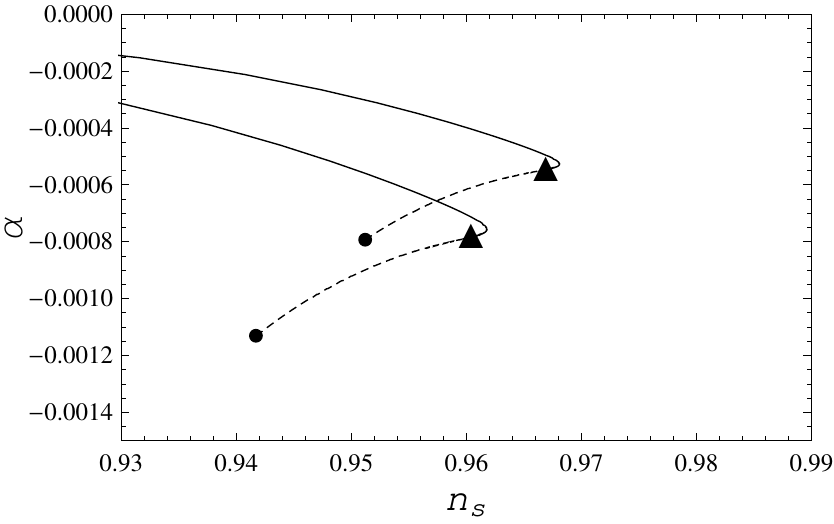}}
\end{center}\vspace{-0.5cm}
\caption{Higgs potential:
$n_s$ vs. $r$ (left panel) and $n_s$ vs. $\alpha$ (right panel)  for various $v$ values, 
 along with the $n_s$ vs. $r$ contours (at the confidence levels of 68\% and 95\%) 
 given by the Planck collaboration (Planck TT+lowP) \cite{Planck:2015xua}. The dashed portions are for $\phi>v$.
The black points and triangles are predictions 
 in the textbook quartic and quadratic potential models, respectively.
$N$ is taken as 50 (left curves) and 60 (right curves).
}
\label{fig:Higgs}
\end{figure}

In the inflationary scenario with the Higgs potential, 
 we can consider two cases for the inflaton VEV during inflation. 
One is that the initial inflaton VEV is smaller than 
 its VEV at the potential minimum ($\phi_0 < v$), 
 and the other is the case with $\phi_0 > v$. For each case, 
we calculate the inflationary predictions for various values 
 of the inflaton VEV keeping the number of e-folds fixed.
Numerical results are displayed in Table \ref{higgstab}.
Figure~\ref{fig:Higgs} shows the predictions for 
 $n_s$, $r$ and $\alpha$ with the number of e-folds 
 $N=50$ (left curves in each panel) and $N=60$ (right curves in each panel).

For the case with $\phi_0 < v$, if the inflaton VEV is large ($v \gg 1$ in
Planck units) the inflation potential is dominated by the VEV term and
well approximated as the quadratic potential, 
\begin{eqnarray} 
V \simeq \left(\frac{\lambda v^2}{6}  \right) \chi^2\,,  
\end{eqnarray} 

\clearpage

\begin{table}[!t]
\begin{center}
\begin{tabular}{cccccccc}
\hline \hline
 $\log_{10}(\lambda)$ & 
 $v$ & 
 $V(\phi_0)^{1/4}$ (GeV)&
 $\phi_0$ &
 $\phi_e$ & 
 $n_s$  & 
 $r$ & 
 $-\alpha\ (10^{-4})$ \\
\hline \hline
\multicolumn{8}{c}{solutions below the VEV ($\phi<v$)}\\\hline
 $-$12.3  & 13  & $1.18 \times 10^{16}$ & 1.91  & 11.7
 & 0.947  & 0.0170  & 2.67  \\\hline
 $-$12.4  & 17  & $1.45 \times 10^{16}$ & 4.64  & 15.7  & 0.960  & 0.0385  & 4.06  \\\hline
 $-$12.6  & 23  & $1.64 \times 10^{16}$ & 9.64  & 21.7  & 0.966  & 0.0626  & 4.82  \\\hline
 $-$12.9  & 32  & $1.76 \times 10^{16}$ & 17.9  & 30.6  & 0.968  & 0.0834  & 5.14  \\\hline
 $-$13.3  & 53  & $1.86 \times 10^{16}$ & 38.3  & 51.6  & 0.968  & 0.104  & 5.32  \\\hline
 $-$14.9  & 300  & $1.96 \times 10^{16}$ & 285  & 299  & 0.967  & 0.128  & 5.44 
\\\hline 
\hline
\multicolumn{8}{c}{solutions above the VEV ($\phi>v$)}\\
\hline
$-12.1$& 
$1$&
$2.33\times 10^{16}$ &  
$22.3$& 
$3.69$&
$0.952$&
$0.258$&
$7.85$\\\hline
$-12.2$&
$5$&
$2.28\times 10^{16}$ & 
$24.3$&
$6.81$& 
$0.955$&
$0.237$&
$7.02$\\\hline 
$-12.5$&
$10$&
$2.22\times 10^{16}$ & 
$28.1$&
$11.6$&
$0.959$&
$0.212$&
$6.36$\\\hline
$-12.8$&
$19$&
$2.15\times 10^{16}$ & 
$36.2$&
$20.5$&
$0.962$&
$0.186$&
$5.91$\\\hline
$-13.3$&
$41$&
$2.08\times 10^{16}$ & 
$57.4$&
$42.5$&
$0.965$&
$0.161$&
$5.65$\\
\hline
$-14.9$&
$300$&
$1.99\times 10^{16}$ & 
$316$&
$301$&
$0.967$&
$0.137$&
$5.49$\\
\hline
\hline
\multicolumn{8}{c}{$V=(1/2) m^2 \phi^2 $}\\\hline 
   &
   &
$1.97 \times 10^{16}$ & 
$15.6$&
$1.41$&
$0.967$&
$0.132$&
$5.46$\\
\hline
\hline
\multicolumn{8}{c}{$V=(1/4!) \lambda \phi^4 $}\\\hline 
$-12.1$&
   &
$2.34\times 10^{16}$ & 
$22.2$&
$3.46$&
$0.951$&
$0.260$&
$7.93$\\
\hline
\hline
\end{tabular}
\end{center}
\caption{Higgs potential:
The values of parameters  
 for number of e-folds $N=60$, in units $m_P=1$ 
 unless otherwise stated. 
} \label{higgstab}
\end{table}

\clearpage

\noindent where $\chi = \phi - v$ plays the role of inflaton.
Thus the predictions approach the values given by \eq{quadratic}, corresponding to the black triangles in Figure~\ref{fig:Higgs}. On the other
hand, for $v\ll1$, the potential is of the new inflation or hilltop type:
\begin{eqnarray} 
V \simeq\frac{\lambda}{4!} v^4\left[1-2\left(\frac{\phi}{v}\right)^2\right] \,,  
\end{eqnarray} 
which implies a strongly red tilted spectrum with suppressed $r$.

In the other case with $\phi_0 > v$, 
 the inflationary predictions for various values of $v$ 
 are shown as dashed lines in Figure~\ref{fig:Higgs}. 
For a small VEV ($v \ll 1$) and $\phi_0 \gg v$, 
 the inflaton potential is well approximated as 
 the quartic potential, and hence the predictions 
 are well approximated by \eq{quartic}, corresponding to the 
 black points in Figure~\ref{fig:Higgs}.
On the other hand, for $v \gg1$ the potential during the observable part of
inflation is approximately the quadratic potential, so that 
 the inflationary predictions approach the values given by \eq{quadratic}
 as the inflaton VEV is increased.

\section{Coleman-Weinberg potential}
In this section we briefly review a class of models which appeared in the 
early eighties in the framework of non-supersymmetric GUTs and employed a GUT singlet scalar field
$\phi$ \cite{Shafi:1983bd,Pi:1984pv,Shafi:1984tt}.
These (Shafi-Vilenkin) models are based on a
Coleman-Weinberg potential \cite{Coleman:1973jx}
which can be expressed as
\cite{Albrecht:1984qt}:
\begin{equation} \label{potpot}
                  V(\phi)= A \phi^4 \left[\ln\left( \frac{\phi}{v}\right)
-\frac{1}{4}\right] + \frac{A v^4}{4}\,,
\end{equation}
where $v$ denotes the $\phi$ VEV at the minimum. Note that $V(\phi=v)=0$,
and the vacuum energy density at the origin is given by $V_0 = A v^4 /4$.
Inflationary predictions of this potential was recently analyzed in
refs. \cite{Smith:2008pf,Rehman:2008qs,Shafi:2006cs}.

The magnitude of $A$ and the inflationary parameters can be calculated using the
standard slow-roll expressions given in \sektion{intro}. 
For $V_0^{1/4}\gtrsim2\times10^{16}$ GeV, observable inflation occurs close to the minimum where the potential is effectively quadratic as in
\sektion{higgs} ($V\simeq2Av^2 \chi^2$, where $\chi=\phi-v$ denotes the
deviation of the field from the minimum).
The inflationary predictions are thus approximately given by \eq{quadratic}.

For $V_0^{1/4}\lesssim10^{16}$ GeV, assuming inflation takes place with inflaton
values below $v$, the inflationary parameters are
similar to those for new inflation models with $V=V_0[1-(\phi/\mu)^4]$:
$n_s\simeq1-(3/N)$,
$\alpha\simeq -3/N^2$. We also consider the case where
inflation takes place at inflaton values above $v$ (see also \cite{Rehman:2008qs}), in which case for 
$V_0^{1/4}\lesssim10^{16}$ GeV
the inflationary parameters are similar to those for the quartic potential given
by \eq{quartic}.

\begin{figure}[!t]
\begin{center}
\scalebox{0.67}{\includegraphics{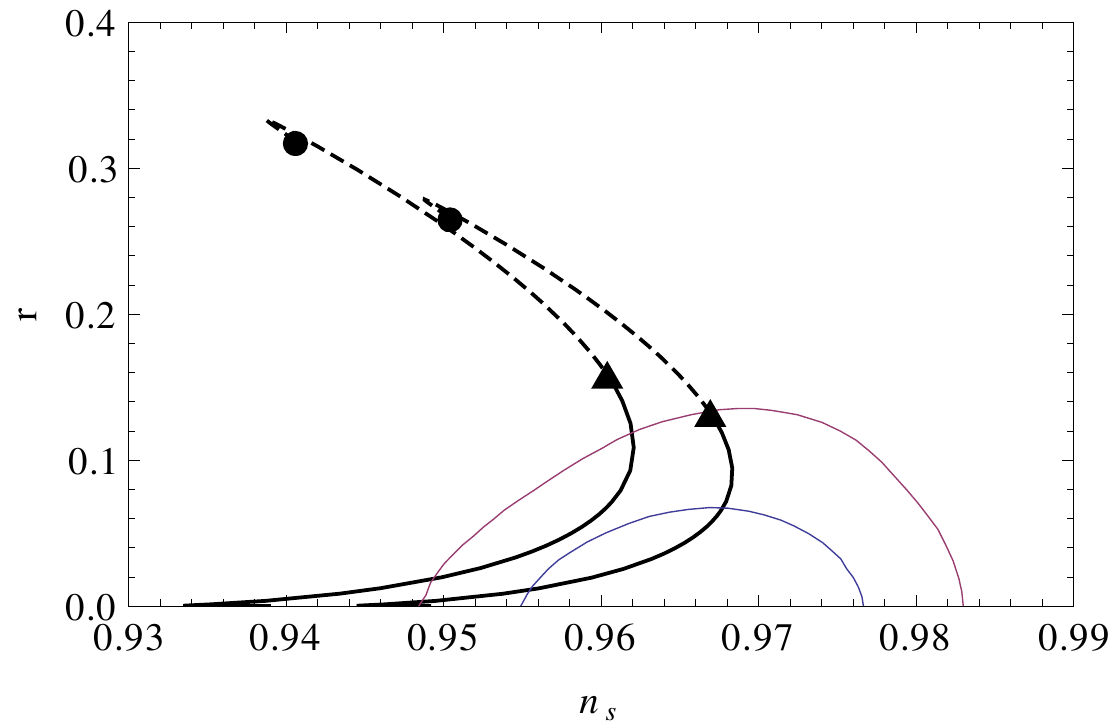}}
\scalebox{0.67}{\includegraphics{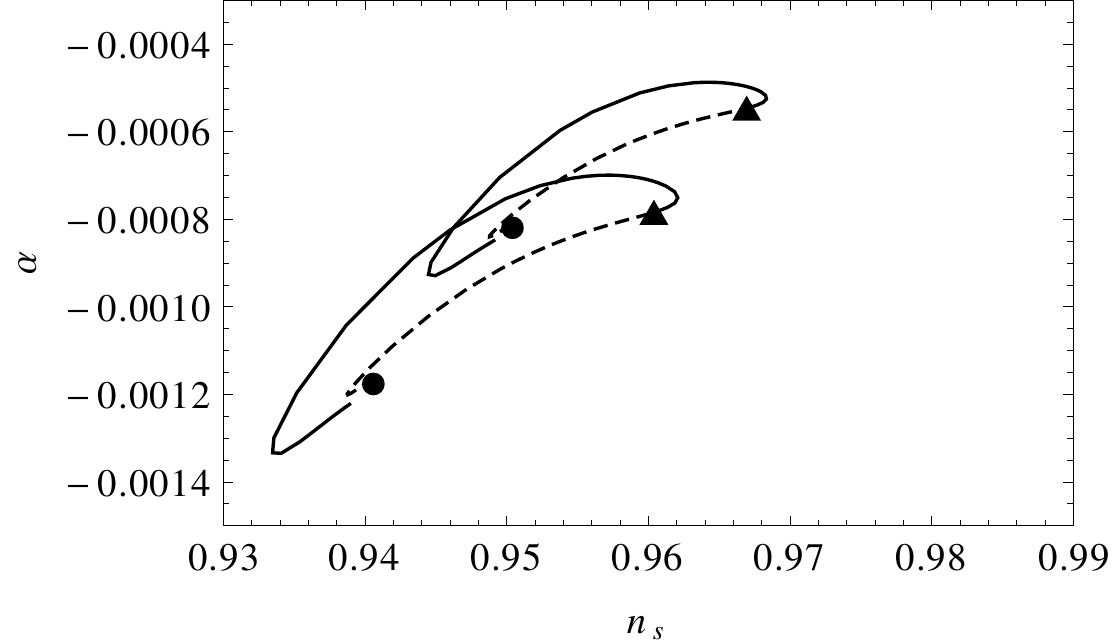}}
\end{center}\vspace{-0.5cm}
\caption{Coleman-Weinberg potential:
$n_s$ vs. $r$ (left panel) and $n_s$ vs. $\alpha$ (right panel) for various $v$ values, 
along with the $n_s$ vs. $r$ contours (at the confidence levels of 68\% and 95\%) 
 given by the Planck collaboration (Planck TT+lowP) \cite{Planck:2015xua}.  The dashed portions are for $\phi>v$.
The black points and triangles are predictions 
 in the textbook quartic and quadratic potential models, respectively.
$N$ is taken as 50 (left curves) and 60 (right curves). 
}
  \label{cwfig}
\end{figure}

\begin{figure}[!th]
\centering
\includegraphics[angle=0, width=8cm]{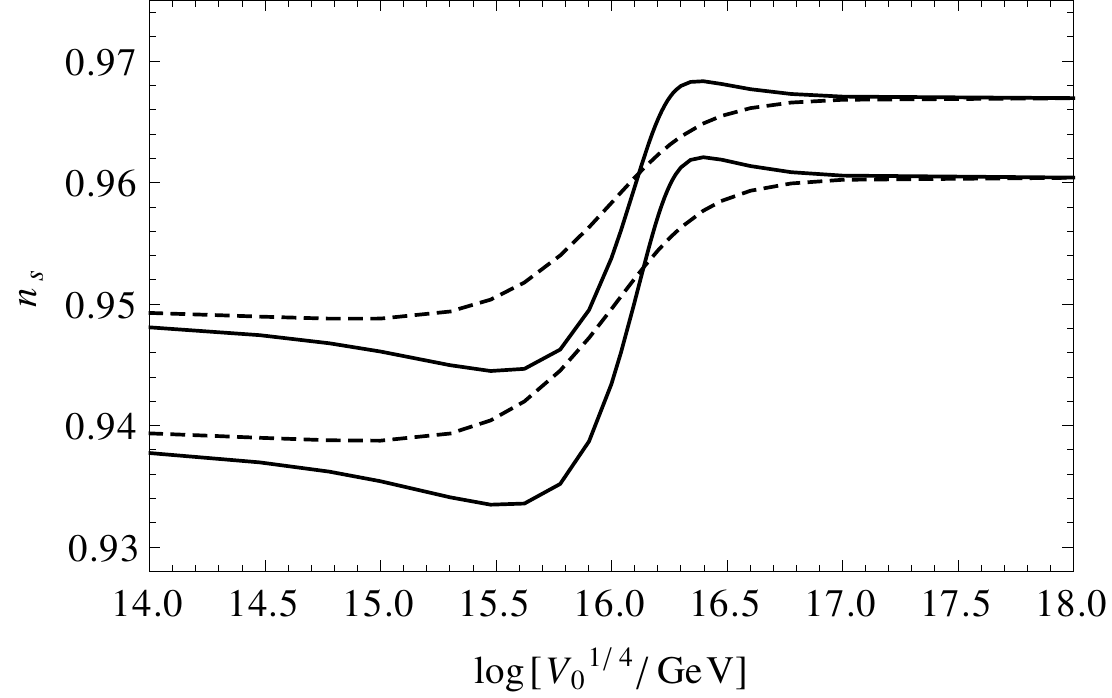}
  \caption{$n_s$ vs. $\log[V_0^{1/4}/\rm{GeV}]$ for the Coleman-Weinberg
potential. The dashed portions are for $\phi>v$. Top to bottom: $N=60,50$.  }
  \label{cwfig2}
\end{figure}

We display the predictions for $n_s$, $r$ and $\alpha$ in \fig{cwfig}. The
dependence of $n_s$ on $V_0$ is displayed in \fig{cwfig2}. Numerical results for
selected values of $V_0$ are displayed in Table \ref{cwtab}. Note that in the
context of non-supersymmetric GUTs, $V_0^{1/4}$ is related to the unification
scale, and is typically a factor of 3--4 smaller than the superheavy gauge boson
masses due to the loop factor in the Coleman-Weinberg potential.  For a discussion of inflation in non-supersymmetric GUTs such as $SU(5)$ and
$SO(10)$ with a unification scale of order $10^{16}$ GeV, see \ocite{Rehman:2008qs}.
As discussed in \ocite{Shafi:1984tt}, in this class of models it is possible for
cosmic topological defects to survive inflation, remaining at an observable
level. 

\begin{table}[!t]
\begin{center}
\begin{tabular}{ccccccccc}
\hline \hline
 {\small $V_0^{1/4}$(GeV)} &  {\small $V(\phi_0)^{1/4}$(GeV)} & {\small
$A(10^{-14})$} & $v$ & $\phi_0$ &
$\phi_e$ & $n_s$  & $r$ & {\small $-\alpha(10^{-4})$}    \\
\hline \hline
\multicolumn{9}{c}{solutions below the VEV ($\phi<v$)}\\\hline
 $1.\times 10^{15} $& $1.\times 10^{15}$ & 1.60 & 1.63 & 0.034 & 0.898 & 0.946 &
 $10^{-6}$ & 9.11 \\\hline
 $1.\times 10^{16} $& $9.92\times 10^{15}$ & 4.37 & 12.7 & 3.38 & 11.4 & 0.954 & 0.008 & 5.97 \\\hline
 $1.5\times 10^{16}$ &$ 1.43\times 10^{16}$ & 2.41 & 22.1 & 10.2 & 20.8 & 0.964
& 0.036 & 4.87 \\\hline
 $1.75\times 10^{16}$ &$ 1.58\times 10^{16}$ & 1.43 & 29.4 & 16.5 & 28.0 & 0.967
& 0.055 & 4.95 \\\hline
 $2.\times 10^{16} $& $1.7\times 10^{16}$ & 0.812 & 38.7 & 25.1 & 37.3 & 0.968 &
0.072 & 5.09 \\\hline
 $3.\times 10^{16} $& $1.87\times 10^{16}$ & 0.121 & 93.4 & 78.6 & 92.0 & 0.968 & 0.107 & 5.33 \\\hline
 $6.\times 10^{16} $& $1.95\times 10^{16}$ & 0.0059 & 397. & 382. & 396. &
0.967 & 0.126 & 5.43\\\hline
\hline
\multicolumn{9}{c}{solutions above the VEV ($\phi>v$)}\\\hline
 $6.\times 10^{16}$ & $2.00\times 10^{16}$ & 0.0050 & 414. & 430. & 416. & 0.967 &
0.138 & 5.49 \\\hline
$ 3.\times 10^{16} $& $2.05\times 10^{16}$ & 0.0623 & 110. & 126. & 112. & 0.965 & 0.152 & 5.57 \\\hline
$ 2.\times 10^{16} $& $2.11\times 10^{16}$ & 0.215 & 53.9 & 70.6 & 55.4 & 0.964
& 0.171 & 5.70 \\\hline
$ 1.4\times 10^{16} $&$ 2.17\times 10^{16}$ & 0.496 & 30.6 & 48.0 & 32.2 & 0.961 & 0.193 & 5.93 \\\hline
$ 1.\times 10^{16} $& $2.24\times 10^{16} $& 0.847 & 19.1 & 37.3 & 20.7 & 0.958
& 0.217 & 6.30 \\\hline
$ 6.\times 10^{15} $& $2.31\times 10^{16} $& 1.29 & 10.3 & 29.7 & 12.1 & 0.954 & 0.247 & 7.02 \\\hline
$ 1.\times 10^{15} $& $2.38\times 10^{16} $& 1.20 & 1.76 & 23.8 & 4.64 & 0.949 & 0.276 & 8.24 \\\hline
$ 1.\times 10^{13} $& $2.36\times 10^{16} $& 0.50 & 0.022 & 22.6 & 3.67 & 0.950
& 0.269 & 8.10\\\hline
\hline
\end{tabular}
\end{center}
\caption{Coleman-Weinberg potential:
The values of parameters  
 for number of e-folds $N=60$, in units $m_P=1$ 
 unless otherwise stated. 
} \label{cwtab}
\end{table}

\section{Quartic potential with non-minimal gravitational coupling} \label{nonm}
Finally we consider a quartic inflaton potential 
 with a non-minimal gravitational coupling~\cite{Okada:2010jf,Okada:2011en,
Salopek:1988qh}. 
One of the simplest scenarios of this kind 
 is the so-called Higgs inflation, 
 which has received a fair amount of
attention~\cite{Bezrukov:2007ep}.
In  Higgs inflation, the SM Higgs field plays 
 the role of inflaton with a strong non-minimal
 gravitational interaction and a typical prediction is 
 $(n_s, r)=(0.968, 0.003)$ for  $N=60$ e-folds.\footnote{The predictions of
SM Higgs inflation depend sensitively on the Higgs and top quark masses, and a
larger $r$ value is also possible, see ref.
\cite{Hamada:2014iga} and references therein. For SM Higgs inflation with a
non-minimal coupling of the kinetic term, see \ocite{Germani:2014hqa}.}  
In non-minimal $\phi^4$ inflation, 
 the inflationary predictions vary from 
 those in   $\phi^4$ inflation (\eq{quartic}) 
 to those in  Higgs inflation, depending 
 on the strength of the non-gravitational
coupling~\cite{Martin:2014vha,Okada:2010jf,Okada:2011en,Salopek:1988qh}.
Non-minimal $\phi^4$ inflation can be embedded 
 into well-motivated particle physics
models~\cite{Okada:2011en,Lerner:2009xg}. Radiative corrections to the potential
have been considered in refs. \cite{Martin:2014vha,Okada:2010jf,Okada:2011en}.

The basic action of  non-minimal $\phi^4$ inflation 
 is given in the Jordan frame 
\begin{eqnarray}
S_J^{{\rm tree}} = \int d^4 x \sqrt{-g} 
\left[- \left( \frac{1+ \xi \phi^2}{2}\right)\mathcal{R}
+\frac{1}{2} (\partial \phi)^2 - \frac{\lambda }{4!} \phi^4 \right],
\end{eqnarray}
 where $\phi$ is a gauge singlet scalar field, and 
 $\lambda$ is the self-coupling. 
We rewrite the action in the Einstein frame as 
\begin{eqnarray}
 S_E = \int d^4 x \sqrt{-g_E}\left[ -\frac{1}{2} \mathcal{R}_E+
 \frac{1}{2} (\partial_E \sigma_E)^2 -V_E(\sigma_E(\phi))\right],
\end{eqnarray}
 where the canonically normalized scalar field has a relation to 
 the original scalar field as 
\begin{eqnarray}
\left(\frac{d \sigma}{d \phi}\right)^{-2} =
 \frac{\left( 1 + \xi \phi^2 \right)^2}{1+(6 \xi +1) \xi \phi^2},  
\end{eqnarray}
 and the inflation potential in the Einstein frame is 
\begin{eqnarray}
V_E(\sigma_E(\phi)) = 
 \frac{\frac{1}{4!} \lambda(t) \phi^4}
 {\left(1+\xi\,\phi^2 \right)^2}.
\end{eqnarray}

The inflationary slow-roll parameters in terms of the original 
 scalar field ($\phi$) are expressed as 
\bea
\epsilon(\phi)&=&\frac{1}{2} \left(\frac{V_E'}{V_E \sigma'}\right)^2, 
 \nonumber \\
\eta(\phi)&=& 
\frac{V_E''}{V_E (\sigma')^2}- \frac{V_E'\sigma''}{V_E (\sigma')^3} ,  
 \nonumber \\
\zeta (\phi) &=&  \left(\frac{V_E'}{V_E \sigma'}\right) 
 \left( \frac{V_E'''}{V_E (\sigma')^3}
-3 \frac{V_E'' \sigma''}{V_E (\sigma')^4} 
+ 3 \frac{V_E' (\sigma'')^2}{V_E (\sigma')^5} 
- \frac{V_E' \sigma'''}{V_E (\sigma')^4} \right)  , 
\eea
where a prime denotes a derivative with respect to $\phi$. 
Accordingly, the number of e-folds is given by
\begin{eqnarray}
N=\frac{1}{\sqrt2} \int_{\phi_{\rm e}}^{\phi_0}
\frac{d\phi}{\sqrt{\epsilon(\phi)}}\left(\frac{d\sigma}{d\phi}\right)\, .
\label{Ne}
\end{eqnarray}

Once the non-minimal coupling $\xi$ and the number 
 of e-folds $N$ are fixed, 
 the inflationary predictions for $n_s$, $r$, and $\alpha$ 
 are obtained. 
Approximate formulas for the predictions of  
 non-minimal $\phi^4$ inflation are given by~\cite{Okada:2010jf}:
\begin{eqnarray}
n_s &\simeq&  1 - \frac{3(1+16\,\xi N/3)}{N\,(1+8\,\xi N)}, \\
r &\simeq&  \frac{16}{N\,(1+8\,\xi N)}, \\
\alpha&\simeq& -\frac{3\,\left(1+4\,(8\,\xi N)/3-5\,(8\,\xi N)^2
-2\,(8\,\xi N)^3 \right)}{N^2\,(1+8\,\xi N)^4}
+ \frac{r}{2} \left(\frac{16\,r}{3} -(1 - n_s)\right).
\end{eqnarray}
The predictions in the textbook quartic potential model 
 are modified in the presence of the non-minimal coupling $\xi$. For $\xi>0$, 
these results exhibit a reduction in the value of $r$ 
 and an increase in the value of $n_s$, as $\xi$ is increased. 
%
Here we have varied $\xi$ along each curve 
 from $0$ to $\xi \gg 1$. The numerical results for selected values of $\xi$ are
displayed in Table \ref{nonmintab}.
The predicted values of $n_s$, $r$ and $\alpha$ 
 are shown in Figure~{\ref{fig:nonminimal} 
 for the number of e-folds $N = 50$ (left curves in each panel) 
 and $N = 60$ (right curves in each panel),  
along with the $n_s$ vs. $r$ contours
given by the Planck collaboration \cite{Planck:2015xua}.

\begin{table}[!htbp]
\begin{center}
\begin{tabular}{cccccccc}
\hline \hline
$\xi$ &
$\log_{10}(\lambda)$ & 
$V(\phi_0)^{1/4}$ (GeV)&
$\phi_0$ &
$\phi_e$ & 
$n_s$  & 
$r$ & 
$-\alpha\ (10^{-4})$ \\
\hline \hline
$10^{-5}$&
$-12.1$&
$2.34\times 10^{16}$ & 
$22.2$&
$3.46$&
$0.951$&
$0.259$&
$7.93$\\
\hline
$3.98 \times 10^{-4}$&
$-12.0$&
$2.24\times 10^{16}$ & 
$22.2$&
$3.45$&
$0.954$&
$0.218$&
$7.86$\\
\hline
$0.001$&
$-11.9$&
$2.12\times 10^{16}$ & 
$22.2$&
$3.43$&
$0.957$&
$0.174$&
$7.65$\\
\hline
$0.002$&
$-11.8$&
$1.97\times 10^{16}$ & 
$22.1$&
$3.40$&
$0.959$&
$0.131$&
$7.29$\\
\hline
$0.00398$&
$-11.6$&
$1.79\times 10^{16}$ & 
$22.0$&
$3.34$&
$0.962$&
$0.0884$&
$6.79$\\
\hline 
$0.01$&
$-11.3$&
$1.51\times 10^{16}$ & 
$21.7$&
$3.18$&
$0.965$&
$0.0451$&
$6.12$\\
\hline
$1.00$&
$ -8.55$&
$0.794\times 10^{16}$ & 
$8.52$&
$1.00$ & 
$0.968$&
$0.00346$&
$5.25$\\
\hline
$100$&
$-4.62$&
$0.764\times 10^{16}$ & 
$0.920$&
$0.107$&
$0.968$&
$0.00297$&
$5.23$\\
\hline
\hline
\multicolumn{8}{c}{$V=(1/4!) \lambda \phi^4 $}\\\hline 
   &
$-12.1$&
$2.34\times 10^{16}$ & 
$22.2$&
$3.46$&
$0.951$&
$0.260$&
$7.93$\\
\hline
\hline
\end{tabular}
\end{center}
\caption{$\phi^4$ potential with non-minimal gravitational coupling:
The values of parameters 
for number of e-folds $N=60$, in units $m_P=1$ 
 unless otherwise stated. 
} \label{nonmintab}
\end{table}

\begin{figure}[!htbp]
\begin{center}
\scalebox{0.4}{\includegraphics{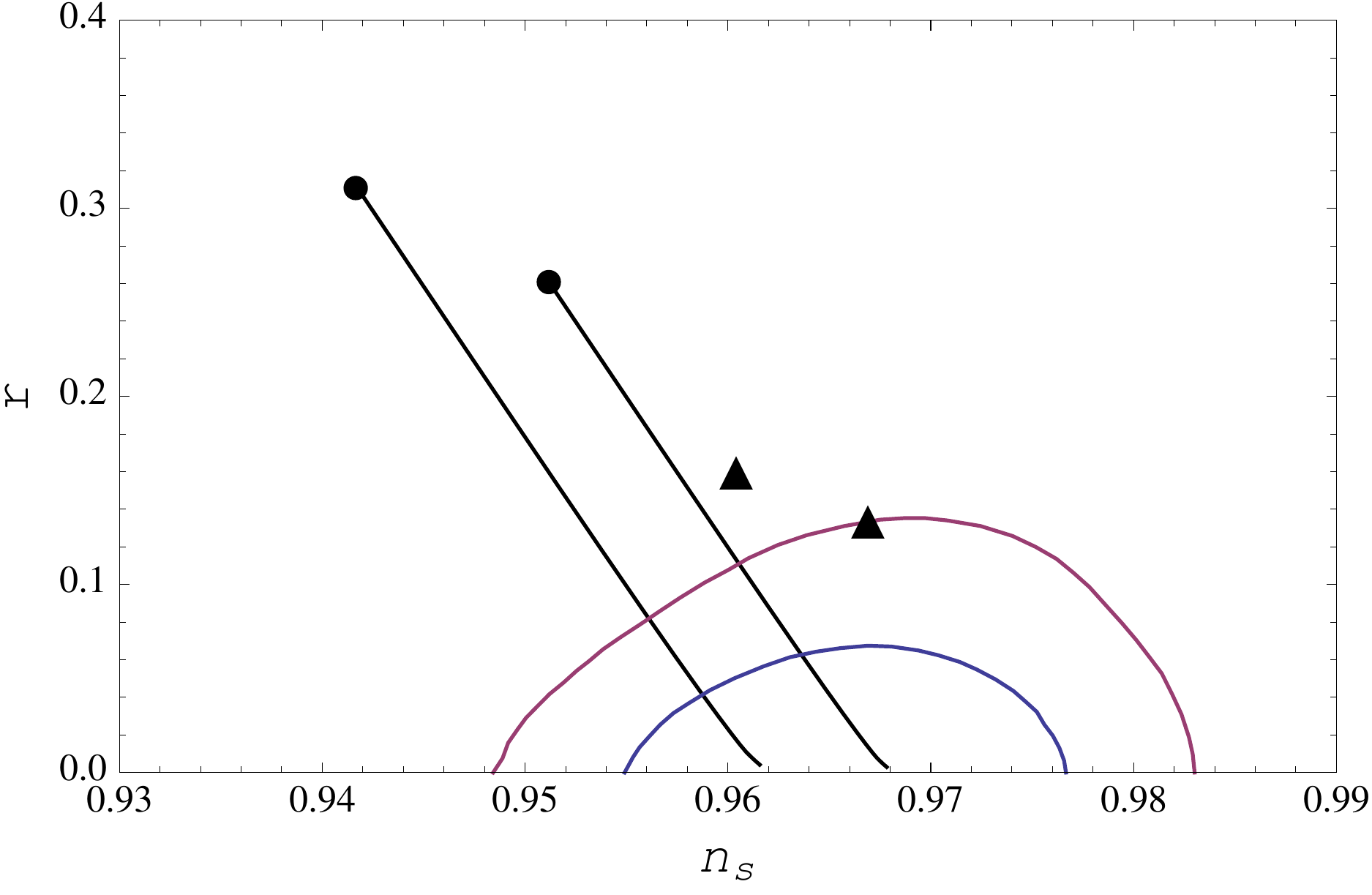}}
\scalebox{0.9}{\includegraphics{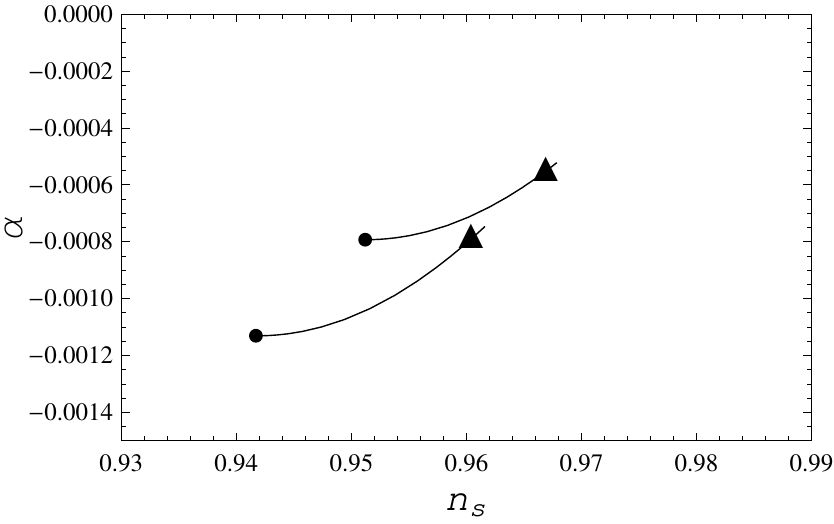}}
\end{center}\vspace{-0.5cm}
\caption{$\phi^4$ potential with non-minimal gravitational coupling:
$n_s$ vs. $r$ (left panel) and $n_s$ vs. $\alpha$ (right panel) for various $\xi$ values, 
along with the $n_s$ vs. $r$ contours (at the confidence levels of 68\% and 95\%) 
 given by the Planck collaboration (Planck TT+lowP) \cite{Planck:2015xua}. 
The black points and triangles are predictions 
 in the textbook quartic and quadratic potential models, respectively. 
$N$ is taken as 50 (left curves) and 60 (right curves).
}
\label{fig:nonminimal}
\end{figure}

\section{Conclusion}

We have restricted our attention in this paper to
models based on relatively simple non-supersymmetric inflationary potentials
involving a SM (or even GUT) singlet scalar field. In the framework of slow-roll
inflation,  a tensor to scalar ratio $r \sim 0.02$--0.1 for spectral index $n_s\simeq0.96$ 
is readily obtained in
these well motivated models. This range of $r$ is of great interest as it is 
experimentally accessible in the very near future. The running of the spectral index in all these models is predicted to be fairly
small, $|\alpha|$ being of order few$\times 10^{-4}$--$10^{-3}$.

For the Higgs and Coleman-Weinberg
potentials, a more precise measurement of $r$ should enable one to ascertain whether the
inflaton field was larger or smaller than its  VEV during the last 60
or so e-folds (the current data favors the latter). 
For the quadratic and quartic inflationary potentials we have
emphasized, following earlier work, that the well-known predictions for $n_s$
and $r$ can be significantly altered if the inflaton couplings to additional
fields, necessarily required for reheating, are taken into account. Despite
these radiative corrections, the predictions for the quartic potential are not
compatible with the current data. A more precise determination of $n_s$ and $r$ 
should enable one to also test the radiatively corrected quadratic model.

We also explored inflation driven by a
quartic potential with an additional non-minimal coupling of the inflaton field
to gravity. With plausible values for the new dimensionless parameter $\xi$
associated with this coupling, the predictions for $n_s$ and $r$ are in good
agreement with the observations. 


\section*{Acknowledgements} Q.S. acknowledges support provided by the DOE grant
No. DE-FG02-12ER41808. We thank Mansoor Ur Rehman for useful comments on the
draft version.

\bibliographystyle{utphys.bst} 
\bibliography{simple3}

\end{document}